\documentclass[twocolumn,showpacs,preprintnumbers,amsmath,amssymb,superscriptaddress,showkeys]{revtex4}

\usepackage{graphicx}
\usepackage{dcolumn}
\usepackage{bm}
\usepackage{enumerate}
\newcommand{\pom}{I\!\! P}
\def\met{\not\!\!E_T}
\input{psfig.sty}
\begin{document}

\hyphenation{PYTHIA DD SD ND}

\title{Diffractive cross sections and event final states at the LHC \footnote{ Presented at {\em Forward Physics at LHC Workshop (May 27-29 2010),} Elba Island, Italy.}}

\author{Konstantin Goulianos}

\affiliation{The Rockefeller University}

\date{\today}


\begin{abstract}
We discuss a phenomenological model that describes results on diffractive $pp$ and $\bar pp$ cross sections and event final states up to the Fermilab Tevatron energy of $\sqrt s =1.96$ TeV and use it to make predictions for Large Hadron Collider (LHC) energies up to $\sqrt s=14$~TeV and asymptotically as $\sqrt s\rightarrow\infty$. 
 The model is anchored in a saturation effect observed in single diffraction dissociation that explains quantitatively the factorization breaking observed in soft and hard $pp$ and $\bar pp$ diffractive processes and in diffractive photoproduction and low $Q^2$ deep inelastic scattering.    
\end{abstract}
\maketitle

\section{Introduction}

As we entered a new energy frontier at the Large Hadron Collider (LHC) with data collected at $\sqrt s=900$~GeV, 2360~GeV, and 7~TeV from Fall 2009 to Spring 2010, it became painfully clear that the Monte Carlo (MC) simulations designed to represent the collective knowledge of the field on diffractive cross sections and event final states did not meet the challenge presented to them in this new higher energy environment. The most commonly used event generators, {\sc pythia}~\cite{PYTHIA} and {\sc phojet}~\cite{PHOJET}, were found to disagree not only with the data but also with each other. The latter clearly meant that the two simulations could not both be right. Therefore, an update of the MCs was urgently needed. Because of the importance of Minimum-Bias (MB) MC simulations in estimating trigger rates, backgrounds, and the machine luminosity at the LHC, a ``{\sc diffraction}'' workshop was organized at CERN on 7 May 2010~\cite{D-day} that brought experimentalists and theorists together to exchange ideas with the goal of producing a reliable MC generator for the LHC. This paper is based on a talk I presented at that meeting and an expanded version presented at this workshop.  
  
Diffraction dissociation in $pp$/$\bar pp$ interactions may be defined by the signature of one or more ``large'' and characteristically not exponentially suppressed~\cite{Bj} rapidity gaps (regions of rapidity devoid of particles)~\cite{rapidity} in the final state. The rapidity gap is presumed to be due to the exchange of a strongly-interacting color singlet quark/gluon combination with the quantum numbers of the vacuum, traditionally referred to as ``Pomeron''~($\pom$). Diffractive processes are classified as single diffraction (SD), double Pomeron exchange (DPE), also referred to as central dissociation (CD), and double diffraction (DD). In $\bar pp$ SD$_{\bar p}$~(SD$_p$), the $p(\bar{p})$ dissociates while the $\bar{p}(p)$ remains intact escaping the collision with momentum close to that of the original beam momentum and separated from the $p$~($\bar p$) dissociation products by a {\em forward} gap; in DPE both the $\bar p$ and the $p$ escape, resulting in {\em two forward gaps}; and in DD a {\em central} gap is formed while both the $p$ and $\bar p$ dissociate. 
The above basic diffractive processes are listed below, along with two additional two 2-gap processes which are combinations of SD and DD and are indicated as SDD:

\begin{figure}[!tb]
\centerline{\psfig{figure=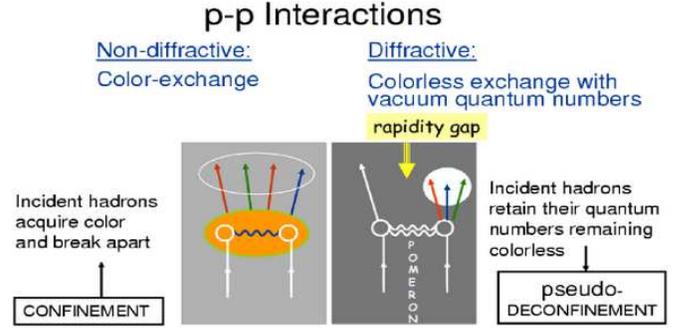,width=0.5\textwidth}}    
\caption{Non-diffractive and diffractive $pp$ interactions.}
\label{fig:rapgaps}
\end{figure}

\begin{table}[h]
\label{cross_sections}
\begin{center}
\caption{Diffractive cross sections.}
\begin{tabular}{ll}\hline\hline
\underline{acronym}&\underline{basic diffractive processes}\\
{\bf SD}$_{\bar p}$&$\bar{p}p\rightarrow \bar{p}+{\rm gap}+[p\rightarrow X_p]$,\\
{\bf SD}$_p$&$\bar{p}p\rightarrow [\bar p\rightarrow X_{\bar p}]+{\rm gap}+p,$\\
{\bf DD}&$\bar{p}p\rightarrow [\bar p\rightarrow X_{\bar p}]+{\rm gap}+[p\rightarrow X_p],$\\
{\bf DPE}&$\bar{p}p\rightarrow \bar{p}+{\rm gap}+X_c+{\rm gap}+p$,\\
&\underline{2-gap combinations of SD and DD}\\
{\bf SDD}$_{\bar p}$&$\bar{p}p\rightarrow \bar{p}+{\rm gap}+X_c+{\rm gap}+[p\rightarrow X_p]$,\\
{\bf SDD}$_p$&$\bar{p}p\rightarrow[\bar{p}\rightarrow X_{\bar p}]{\rm gap}+X_c+{\rm gap}+p$.\\
 \hline\hline
\end{tabular}
\end{center}
\end{table}

\noindent Here, $X_{\bar p}$, $X_p$ and  $X_c$ represent clusters of particles in rapidity regions not occupied by the gap(s). The 2-gap processes are examples of {\em multi-gap} diffraction, a term coined by this author to represent events with multiple diffractive rapidity gaps. A special case of DPE  is {\em  exclusive production,} where a particle state is centrally produced, as for example a dijet system or a $Z$ boson.
 
\begin{figure*}[t]
\unitlength 0.95in
\thicklines
\begin{center}
\begin{picture}(6,1)(1,0)
\put(1,0){\line(2,0){6}}
\multiput(2,0)(2,0){3}{\oval(0.8,0.5)[t]}
\put(1.9,-0.25){$\eta_1'$}
\put(2.9,-0.25){$\eta_2$}
\put(3.9,-0.25){$\eta_2'$}
\put(4.9,-0.25){$\eta_3$}
\put(5.9,-0.25){$\eta_3'$}
\put(2.8,0.5){$\Delta \eta_2$}
\put(4.8,0.5){$\Delta \eta_3$}
\put(1.25,-0.5){$t_1$}
\put(2.9,-0.5){$t_2$}
\put(4.9,-0.5){$t_3$}
\put(6.7,-0.5){$t_4$}
\put(1.15,0.5){$\Delta \eta_1$}
\put(1.85,0.5){$\Delta \eta'_1$}
\put(3.85,0.5){$\Delta \eta'_2$}
\put(5.85,0.5){$\Delta \eta'_3$}
\put(6.6,0.5){$\Delta \eta_4$}
\end{picture}
\end{center}
\vspace{3em}
\caption{Average multiplicity $dN/d\eta$ (vertical axis) vs. $\eta$ (horizontal axis) for a process with four rapidity gaps, $\Delta \eta_{i}(i=1-4)$.} 
\label{fig:soft}
\end{figure*}
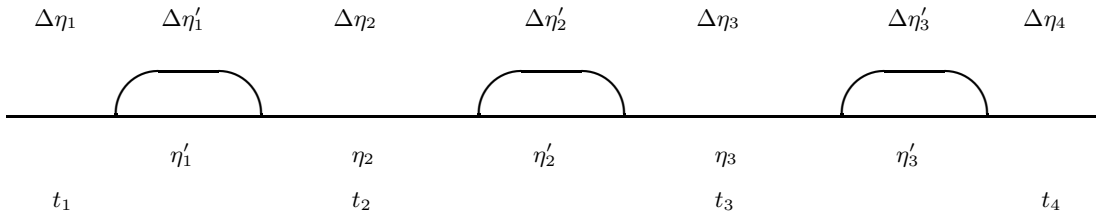

Below, in Sec.~\ref{strategy} ({\sc strategy}) we outline the method we follow to implement an algorithm for a MC simulation, in Sec.~\ref{references} ({\sc cross sections and final states}) we present excerpts from previous papers on total and differential diffractive cross sections and final states, and in Sec.~\ref{conclusions} we conclude.

\section{Strategy}\label{strategy}
A phenomenology that is used to make predictions for the LHC and beyond should be based on cross sections and final states that incorporate the current knowledge in the field molded into a form that can be extrapolated to higher energies. The issues to be addressed is how to take into account saturation effects that suppress cross sections, and what formulas to use for event final state multiplicity, pseudorapidity, and transverse energy ($E_T$)~\cite{missingET} distributions.  In addition, the structure of an algorithm for implementing this knowledge into a MC simulation should also be addressed. The algorithm must be robust against changes in the collision energy, so that it may be equally well applied to simulate collisions at fixed target energies as well as at the higher energies of $pp$ and $\bar pp$ colliders and in astrophysics. In this section, we outline a strategy that addresses these issues.\\

The following input is used for cross sections and final states:

\begin{enumerate}[(i)]
\item $d^2\sigma/d\xi dt$ of the diffractive processes listed in Table~\ref{cross_sections} from the {\sc renorm} model~\cite{R};
\item $\sigma_t(s)$ from {\sc superball model}~\cite{superball};
\item optical theorem $\rightarrow\;Im\,f_{el}(t=0)$ (imaginary part of the forward scattering amplitude); 
\item dispersion relations $\rightarrow\;Re\,f_{el}(t=0)$, using low energy cross sections from global fit~\cite{global};
\item final states: use ``nesting'' to describe gap processes, where a {\em nest} is defined as a region of $\Delta\eta$ where there is particle production, in contrast to a {\em gap} region where there are no particles~\cite{{corfu01},{multigap}}.
\end{enumerate}

 Figure~\ref{fig:soft} shows a schematic $\eta$ topology of an event with four rapidity gaps and three nests of final state particles. The cross section for this configuration is presented in Sec.~\ref{x-sections}

We propose the following algorithm for generating final states:
\begin{itemize}
\item start with a $pp\rightarrow X$ inelastic collision at $\sqrt s$;
\item decide whether the collision is ND or diffractive based on the expected cross sections; if ND, use the ND final state expected at $\sqrt s$; if diffractive, select SD$_{\bar p}$, SD$_p$, DD, or DPE based on probabilities scaled to the corresponding cross sections;
\item for each diffractive event, check whether the region of $\eta$ where particles are produced, $\Delta\eta'$,  is large enough to accommodate additional diffractive rapidity gaps: if {\em yes}, decide whether or not the event will have other gaps within this region, again using probabilities scaled to the cross sections, and branch off accordingly;
\item continue this process until the region $\Delta\eta'$ is too small to accommodate another diffractive gap. 
\end{itemize}

It is important to note that in our definition of a ND collision there are no diffractive gaps whatsoever in the final state of the event. In this respect, this definition differs from those of ``inclusive'' or ``non-single-diffractive'' definitions of ND events used in the literature.     

\section{Cross sections and final states}\label{references} 

In this section,  we discuss briefly the diffraction dissociation and total cross sections using information and/or excerpts from Refs.~\cite{{R},{lathuile04}}.
  
\subsection{Diffractive cross sections}\label{x-sections}
In Ref.~\cite{pomQCD}, the following expression is obtained for the SD cross section [quoting]:

\begin{quote}
\begin{eqnarray}
\frac{d^2\sigma_{sd}(s,\Delta\eta,t)}{dt\,d\Delta\eta}=
\frac{1}{N_{gap}(s)}\times \nonumber\\
\underbrace{C_{gap}\cdot F_p^2(t)\left\{e^{\textstyle (\epsilon+\alpha' \,t)\Delta\eta}\right\}^2}_{\textstyle P_{gap}(\Delta\eta,t)}
 \cdot \;\kappa \cdot \left[\sigma_\circ\,e^{\textstyle \epsilon\Delta\eta'}\right],
\label{eq:diffPM}
\end{eqnarray}
where:\\
(i) the factor in square brackets represents the cross section due to the wee partons in the $\eta$-region of particle production $\Delta\eta'$;\\
(ii)  $\Delta\eta=\ln s$-$\Delta\eta'$ is the rapidity gap;\\
(iii)  $\kappa$ is a QCD color factor selecting color-singlet $gg$ or $q\bar q$ exchanges to form the rapidity gap;\\ 
(iv) $P_{gap}(\Delta\eta,t)$ is a gap probability factor representing the elastic scattering between the dissociated proton (cluster of dissociation particles) and the surviving proton;\\ 
(v)  $N_{gap}(s)$ is the integral of the gap probability distribution over all phase-space in $t$ and $\Delta\eta$;\\ 
(vi)  $F_p(t)$ in $P_{gap}(\Delta\eta,t)$ is the proton form factor $F_p(t)=e^{\displaystyle b_\circ t}$ ... 
; and\\
(vii) $C_{gap}$ is a normalization constant, whose value is rendered irrelevant by the renormalization division by $N_{gap}(s)$. ...\\
By a change of variables from $\Delta\eta$ to $M^2$ using $\Delta\eta'=\ln M^2$ and $\Delta\eta=\ln s-\ln M^2$, Eq.~(\ref{eq:diffPM}) takes the form:
\begin{eqnarray}
\frac{d^2\sigma (s,M^2,t)}{dM^2 dt}=
\left[\frac{\sigma_\circ}{16\pi}\sigma_\circ^{\pom p}\right]
\,\frac{s^{\displaystyle 2\epsilon}}{N(s)}
\;\frac{1}{\left(M^2\right)^{\displaystyle 1+\epsilon}}\;e^{\displaystyle b\,t}\nonumber\\
\;\;\stackrel{\displaystyle s\rightarrow \infty}{\Rightarrow}\;\;
\left[2\alpha' \,e^{\frac{\displaystyle\epsilon\,b_0}{\displaystyle\alpha' }}
\sigma_\circ^{\pom p}\right]
\frac{\ln s^{\displaystyle 2\epsilon}}{\left(M^2\right)^{\displaystyle 1+\epsilon}}\;e^{\displaystyle b\,t},
\label{eq:diffM2}
\end{eqnarray}
where $b=b_0+2\alpha'\ln\frac{\displaystyle s}{M^2}$ [$b$ is the slope of the diffractive $t$-distribution].
Integrating this expression over $M^2$ and $t$ yields the total single diffractive cross section,
\begin{equation}
\sigma_{sd}\stackrel{\displaystyle s\rightarrow \infty}{\rightarrow} 
2\,\sigma_\circ^{\pom p}\;
\exp\left[{\frac{\epsilon\,b_0}{2\alpha '}}\right]\mbox{= constant}\equiv\sigma^{\infty}_{sd}.
\label{eq:sigma_not}
\end{equation}
 
The remarkable property that the total single diffractive cross section becomes constant as $s\rightarrow \infty$ is a direct consequence of the coherence condition required for the recoil proton to escape the interaction intact. This condition selects one out of several available wee partons to provide a color-shield to the exchange and enable the formation of a diffractive rapidity gap.
\end{quote}

Details are presented in Ref.~\cite{pomQCD}, where this formulation of the cross section  is used to derive the ratio of the intercept to the slope of the Pomeron trajectory. Good agreement with the ratio extracted from measurements is obtained, providing support for the renormalization approach used in the phenomenology.  

A similar expression may be use for DD, DPE, and multigap processes, as discussed in Refs.~\cite{{corfu01},{multigap}}. For example, the differential cross section for the process displayed in Fif.~\ref{fig:soft} is derived in Ref.~\cite{multigap} as [quoting]:

\begin{eqnarray}
\frac{d^{10}\sigma^D}{\Pi_{i=1}^{10}dV_i}=
N^{-1}_{gap}\; 
\underbrace{
F^2_p(t_1)F^2_p(t_4)
\Pi_{i=1}^4\left\{e^{[\epsilon+\alpha't_i]\Delta\eta_i}\right\}^2
}_{\hbox{gap probability}}\;\nonumber\\
\times
\kappa^4\left[\sigma_0\,e^{\epsilon\sum_{i=1}^3\Delta\eta'_i}\right],
\label{eq:diffPM4}
\end{eqnarray}

\begin{quote}

\noindent where the term in square brackets is the $pp$  
total cross section at the reduced $s$-value, defined 
through $\ln (s'/s_0)=\sum_i\Delta \eta_i'$, 
$\kappa$ (one for each gap) is the QCD color factor for gap formation, 
the gap probability is the amplitude squared for elastic scattering 
between two diffractive clusters or between a diffractive cluster and a 
surviving proton with form factor $F^2_p(t)$,
and $N_{gap}$ is the (re)normalization factor defined as 
the gap probability integrated 
over all 10 independent variables $t_i$, $\eta_i$, $\eta'_i$, and 
$\Delta\eta\equiv \sum_{i=1}^4\Delta\eta_i$. 

The renormalization 
factor $N_{gap}$ is a function of $s$ only.
The color factors are $c_g=(N_c^2-1)^{-1}$ 
and $c_q=1/N_c$ for gluon and quark color-singlet exchange, respectively.  
Since the reduced energy cross section is properly normalized, the 
gap probability is (re)normalized to unity. The quark to gluon 
fraction, and thereby the Pomeron intercept parameter $\epsilon$ 
may be obtained from the inclusive parton distribution 
functions (PDFs)~\cite{lathuile04}. Thus, normalized differential multigap cross 
sections at $t=0$ may be fully derived from inclusive PDFs and QCD color 
factors without any free parameters.

The exponential dependence of the cross section on $\Delta\eta_i$ leads to 
a renormalization factor $\sim s^{2\epsilon}$ independent of the number of 
gaps in the process. This remarkable property of the renormalization model, 
which was confirmed in two-gap to one-gap cross section ratios measured by 
the CDF Collaboration (see Ref.~\cite{lathuile04}), suggests that multigap
diffraction can be used as a tool for exploring the QCD aspects of 
diffraction in an environment free of rapidity gap suppression effects.
The LHC with its large rapidity coverage provides the ideal arena 
for such studies.  
\end{quote}

\subsection{The total cross section} 
In Ref.~\cite{superball}, an analytic expression is obtained for the total cross section using a parton model approach and exploiting a saturation effect observed in the SD cross section. The abstract of Ref.~\cite{superball} reads [quoting]:
\begin{quote}
  The single-diffractive and total $pp$ cross sections at the LHC are predicted in a phenomenological approach that obeys all unitarity constraints. The approach is based on the renormalization model of diffraction and a saturated Froissart bound for the total cross section yielding $\sigma_t=(\pi/s_o)\cdot \ln^2(s/s_F)$ for $s>s_F$, where the parameters $s_o$ and $s_F$ are experimentally determined from the $\sqrt s$-dependence of the single-diffractive cross section.  
\end{quote}

\begin{figure}
\vspace*{-5em}
\centerline{\psfig{figure=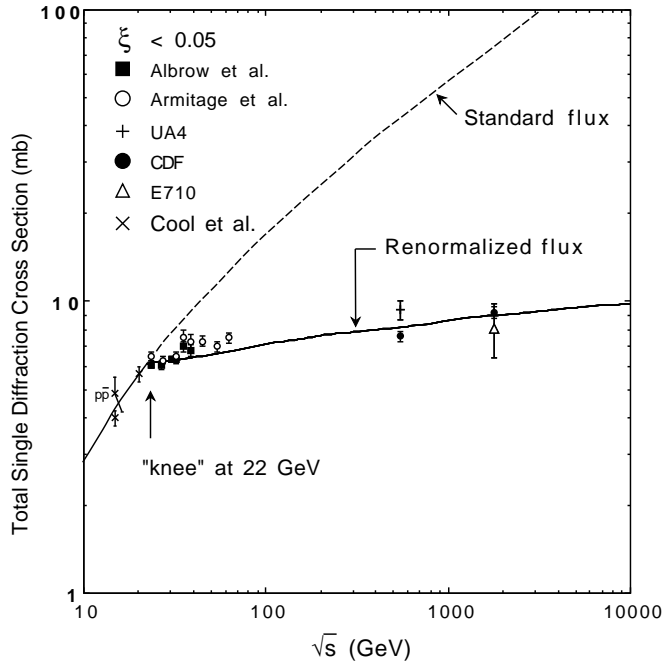,width=0.6\textwidth}}
\vspace*{-10em}  
\caption{Total \protect{$pp/\bar pp$}
single-diffraction dissociation cross section data (sum of both $\bar p$ and $p$ dissociation) for \protect{$\xi<0.05$} compared with predictions 
based on the standard and the renormalized Pomeron flux (from Ref.~\cite{R}).}
\label{fig:tot_sd}
\end{figure}

The following strategy is used in Ref.~\cite{superball} [quoting]:
 \begin{quote}

\addtolength{\itemsep}{-0.5em}
\begin{itemize}       
\item Use the Froissart formula as a {\em saturated}\, cross section rather than as a bound above $s_F$:\\ 

{\Large {$\sigma_t(s>s_F)=\sigma_t(s_F)+ \frac{\pi}{m^2}\cdot \ln^2\frac{s}{s_F}$}}\\

\item This formula should be valid above the {\em knee} in $\sigma_{sd}$ vs. $\sqrt s$ at $\sqrt s_F=22$~GeV (Fig.~\ref{fig:tot_sd}) and therefore valid at $\sqrt s=1800$~GeV.
\item Use $m^2=s_o$  in the Froissart formula multiplied by 1/0.389 to convert it to mb$^{-1}$.
\item Note that contributions from Reggeon exchanges  at $\sqrt s=1800$~GeV are negligible, as can be verified from the global fit of Ref.~\cite{global}.
\item Obtain the total cross section at the LHC:\\

{{
$\sigma_t^{\rm LHC}=\sigma_t^{\rm CDF}+
{\dfrac{\pi}{s_o}}\cdot
\left(\ln^2 \dfrac{s^{\rm LHC}}{s_F}-\ln^2 \dfrac{s^{\rm CDF}}{s_F}\right)$}
}\\
\end{itemize}

For a numerical evaluation of $\sigma^{LHC}$ we use as input the CDF cross section at $\sqrt s=1800$~GeV, $\sigma_t^{CDF}=80.03\pm 2.24$~mb, the Froissart saturation energy $\sqrt s_F=22$~GeV, and the parameter $s_o$. \\
...The resulting prediction for the total cross section at the LHC at $\sqrt s=14\;{\rm TeV}$ is:
$$\sigma^{LHC}_{14\,{\rm TeV}}= (80\pm3)+(29\pm 12)=109\pm12\mbox{\;\; mb}.$$
\end{quote}

\noindent For $\sqrt s=7$~TeV, the predicted cross section is: 
\begin{equation*}
\sigma_{7\;\rm{TeV}}^{\rm LHC}=98\pm 8\;{\rm mb}\left[{\rm at}\;\sqrt s=7\;{\rm TeV}\right],
\end{equation*}
The result for $\sqrt s=14$~TeV is in good agreement with $\sigma_t^{\rm CMG}=114\pm 5\;\rm{mb}$ obtained by the global fit of Ref.~\cite{global}, where the uncertainty was estimated from $\delta\epsilon$ and the  $s^\epsilon$ dependence from which the value of the parameter $s_o$ was obtained. 

\section{Conclusions}\label{conclusions}

We briefly discuss a phenomenological model that describes available results on diffractive $pp$ and $\bar pp$ cross sections and event final states up to the Fermilab Tevatron energy of $\sqrt s =1.96$~TeV  and refer the reader to previous publications for further details. We also outline a procedure to be used to implement the predictions of the model into a Monte Carlo simulation that is robust against changes in the collision energy, so that it may be equally well applied to simulate collisions at fixed target energies as well as at the higher energies of the Tevatron, the LHC, and beyond. The model is anchored in a saturation effect observed in single diffraction dissociation that explains quantitatively the factorization breaking observed in soft and hard $pp$ and $\bar pp$ diffractive processes and in diffractive photoproduction and low $Q^2$ deep inelastic scattering.    

\section{Acknowledgments}
I would like to thank my colleagues at The Rockefeller University and my collaborators at the Collider Detector at Fermilab for  providing the interactive environment in which this work was made possible, and the organizers of the Workshop for putting together a comprehensive program of presentations at the threshold in time of the opening of a new era in particle physics at the Large Hadron Collider.  
 


\begin{thebibliography}{99}
\bibitem{rapidity}
Rapidity, $y=\frac{1}{2}\ln\frac{E+p_L}{E-p_L}$, and pseudorapidity, $\eta=-\ln\tan\frac{\theta}{2}$, are used interchangeably, since in the kinematic regions of interest they are approximately equal.
\bibitem{missingET} Transverse energy is defined as $E_T=E\sin\theta$, and missing $E_T$ as $\met = |\overrightarrow{\met}|$ with  $\overrightarrow{\met}= -\sum_iE_T^i\hat{n_i}$, where $\hat{n_i}$ is a unit vector perpendicular to the beam axis and pointing at the $i^{th}$ calorimeter tower. The sum $E_T$ is defined by $\sum E_T=\sum_iE_T^i$. Both sums are over all calorimeter towers above the set thresholds. The missing $E_T$ significance is defined as $S_{\not E}\equiv{\not\!E_T}/\sqrt{\Sigma{E_T^2}}$
\end{thebibliography}
\end{document}